\documentclass[aps,preprint,amsmath,amssymb,revtex4,nobibnotes]{revtex4}
\usepackage{mathrsfs}
\usepackage{amsfonts}
\def\wtilde#1{\setbox\z@\hbox{$\m@th#1$}%
 \ifdim\wd\z@>\tw@ em\mathaccent"0\msbfam@5D{#1}%
 \else\mathaccent"0365{#1}\fi}
\renewcommand{\thesection}{\arabic{section}}

\newcommand{\ds}{\displaystyle}
\begin{document}

{\large \sf

\vspace*{2cm}
\title{
{\Huge \sf A Soluble Model of "Higgs boson" as a Composite$^{*\dagger}$
\renewcommand{\thefootnote}{{}}\footnote{\normalsize
$*$ to be included in "Selected Papers and a
Glimpse into the life of Gunnar K\"{a}llen", Edited by C. Jarlskog and
A. C. T. Wu.\\
$\dagger~$This research was supported in part by the U.S. Department of Energy (grant no.
DE-FG02-92-ER40699).}}}

\author{
{\large \sf R. Friedberg$^1$ and T. D. Lee$^{1,2}$}\\
{\normalsize \it 1. Physics Department, Columbia University,}
{\normalsize \it New York, New York  10027, USA}\\
{\normalsize \it 2. China Center of Advanced Science and Technology
(CCAST/World Lab.),}
{\normalsize \it P. O. Box 8730, Beijing
100190, China}}
\vspace{1cm}
December 1, 2009
\vspace{1cm}

\begin{abstract}
Higgs boson may turn out to be a composite. The theoretical description of
such a composite is illustrated by an example of a soluble model.
\end{abstract}
\maketitle
\vspace{1cm}


\vspace{1cm}


\newpage
\section*{\Large \sf 1.  Introduction}
\setcounter{section}{1} \setcounter{equation}{0}
The $\sigma$-model [1] is a successful phenomenological theory of low energy
particle physics. Yet, the $0^+$ $\sigma$-particle itself has never
been identified experimentally [2]. One of the possible reasons for
this failure might be that the spin-parity transformation of
$\sigma$ is the same as that of
\begin{equation}
\sigma^2,~\sigma^3,~\cdots~,~\sigma^n, \cdots ~~.
\end{equation}
Thus, what is a $\sigma$-field in the idealized theoretical model
may appear experimentally as a "composite" due to the possible
mixture of (1.1). Today, a major focus of high energy physics is to
search for Higgs boson. It might be that for similar and other reasons, Higgs boson [3]
could also turn out to be a "composite"  [4-8]. Neither the experimental
identification of $\sigma$ nor that of Higgs boson would correspond to
the usual simple theoretical description of a single pole
in the complex energy plane. In this paper, we explore the theoretical
description of such a composite from a more elementary perspective, by
examining the generalization of a soluble model [9]. The structure of
the model is given in Section 2, and its solution in Sections 3 and 4.

This paper is dedicated to the memory of Gunard K\"{a}llen, who (besides
others) has made important contributions [10-15] to the original soluble model.

\section*{\Large \sf 2.  A Generalized Soluble Model}
\setcounter{section}{2} \setcounter{equation}{0}

We generalize the original $V\rightleftarrows N\theta$ model by
retaining the same fixed Fermion states $V$ and $N$, but replacing
the single $\theta (r)$ field by three boson fields $~A(r), ~B(r)$
and $C(r)$. The Hamiltonian $H$ in the new model is given by
\begin{equation}
H = H_0 + H_1 + H_2
\end{equation}
where
\begin{equation}
H_0 = m_0 V^\dagger V + \sum_k (\lambda_k a^\dagger_k a_k
+ \mu_k b^\dagger_k b_k + \nu_k c^\dagger_k c_k ).
\end{equation}

For  convenience, the entire system is enclosed within a sphere of
large radius $R$. The $s$-wave part of the annihilation field operators $~A(r),
~B(r)$ and $C(r)$ are given in terms of their annihilation
operators $a_k, ~b_k$ and $c_k$  by

\begin{eqnarray}
\begin{array}{lll}
~~A(r) & = & \sum_k (4\pi R \lambda_k)^{-\frac{1}{2}} u_k r^{-1} (\sin k r)
   a_k ~~~,\\
~~B(r) & = & \sum_k (4\pi R \mu_k)^{- \frac{1}{2}} v_k r^{-1} (\sin k r)
   b_k \\
\end{array}
\end{eqnarray}
and
\begin{equation*}
\begin{array}{lll}
C(r)  &= &  \sum_k (4\pi R \nu_k)^{- \frac{1}{2}} w_k r^{-1} (\sin k r)
   c_k ~~~~\
\end{array}
\end{equation*}
with
\begin{eqnarray}
\begin{array}{lll}
\lambda_k= (k^2 + \alpha^2)^{\frac{1}{2}},\\
\mu_k= (k^2 + \beta^2)^{\frac{1}{2}},\\
\nu_k= (k^2 + \gamma^2)^{\frac{1}{2}}
\end{array}
\end{eqnarray}
and $\alpha,~\beta,~\gamma$ the masses of bosons $a,~b$ and $c$.
The functions $u_k, ~v_k, ~w_k$ are convergence factors, which may all be
chosen to be 1 for $k < k_{max}$ and $0$ otherwise.
In (2.3) all summations extend over

\begin{equation}
k = n \pi /R
\end{equation}
with $n = 1,2,3,\cdots ~$. At equal time, we have the anti-commutation relations

\begin{equation}
\{V, ~V^\dagger\} = \{N, ~N^\dagger \} = 1
\end{equation}
and the commutation relations
\begin{equation}
[a_k, ~a_{k'}^\dagger] =[b_k, ~b_{k'}^\dagger] = [c_k, ~c_{k'}^\dagger ] = \delta_{kk'}.
\end{equation}
If one wishes, (2.6) can also be changed into commutation relations,
and $V$ and $N$ would then be bosons.

In (2.2), we set the mass of $N$ to be zero, and the "bare" mass of $V$ to be $m_0$.
The interaction Hamiltonians $H_1$ and $H_2$ are given by
\begin{equation}
H_1 = g(V^\dagger N C(0) + N^\dagger V C^\dagger(0))
\end{equation}
and
\begin{equation}
H_2 = f(V^\dagger N B^\dagger(0)A(0)  + N^\dagger V A^\dagger(0)B(0))~.
\end{equation}
The $g$-coupling governs the transition
\begin{equation}
V\rightleftarrows Nc~.
\end{equation}
and the $f$-coupling gives rise to the scattering

\begin{equation}
N a \rightleftarrows Vb~.
\end{equation}
Thus, when $f=0$ the Hamiltonian is identical to that of the
original $V\rightleftarrows N\theta$ model, with $\theta$ replaced by $c$.

Throughout the paper, we assume $V$ to be \underline{unstable} through $V\rightarrow Nc$
when $R\rightarrow \infty$. Since the mass of $N$ is set to be zero in
the model, $V$ is unstable if its physical mass $m$ is larger than $\gamma$, the mass of
$c$; i.e.,

\begin{equation}
m > \gamma~.
\end{equation}
Thus, in a collision of $Na$, beside the elastic scattering
\begin{equation}
N a \rightarrow Na~~,
\end{equation}
we also have the inelastic process
\begin{equation}
N a \rightarrow Vb \rightarrow Nbc
\end{equation}
provided that the total energy $E$ satisfies
\begin{equation}
E > \beta + \gamma~~,
\end{equation}
the threshold energy of the channel $Nbc$.

We shall assume
\begin{equation}
\alpha < \beta + \gamma
\end{equation}
Hence, in the $Na$ channel at low energy when
\begin{equation}
\alpha < E < \beta + \gamma~,
\end{equation}
there is only the elastic scattering (2.13); at higher
energy when $E > \beta + \gamma$, we have both (2.13) and
the inelastic process (2.14).

Consider first the process

\begin{equation}
Nc \rightleftarrows V \rightleftarrows Nc~.
\end{equation}
Denote the corresponding state vector by

\begin{equation}
|Nc~\rangle \propto \left[ V^\dagger + g(4\pi R)^{-\frac{1}{2}}\sum_k \nu_k^{-\frac{1}{2}} k
w_k (E-\nu_k)^{-1} N^\dagger c_k^\dagger \right] |vac~\rangle ~.
\end{equation}

One can readily verify that it satisfies
\begin{equation}
H|Nc~\rangle  = E | Nc~\rangle ~~.
\end{equation}
At a finite $R$, $E$ satisfies  the eigenvalue equation
\begin{equation}
h_R(E) \equiv E - m_0 - g^2 \sum_k {k^2 w_k^2\over 4\pi \nu_k R}
\left({1 \over E - \nu_k }\right) = 0~~~,
\end{equation}
with its derivative
\begin{equation}
h'_R (E) = 1 + g^2 \sum_k {k^2 w^2_k \over 4 \pi \nu_k  R}
\left( {1\over E - \nu_k }\right)^2
\end{equation}
always positive.

When $R \rightarrow \infty$, $h_R(E)$ becomes
\begin{equation}
h_\infty (E) = E - m_0 - g^2 \int_0^\infty {k^2 w_k^2\over 4\pi^2 \nu_k }
{dk \over (E - \nu_k) } ~.
\end{equation}
The condition  for $V$ being unstable is that when $E = \gamma$,
\begin{equation}
h_\infty (\gamma) < 0~~.
\end{equation}
In this case, we introduce a
cut along the real axis from
\begin{equation}
E = \gamma ~~~\hbox{to}~~~\infty
\end{equation}
where $\gamma$ is the mass of the $c$-meson. The derivative of $h_\infty (E)$ is
\begin{equation}
h'_\infty (E) = 1 + g^2 \int_0^\infty {k^2 w_k^2\over 4\pi^2 \nu_k }
{dk \over (E - \nu_k)^2 } ~.
\end{equation}
which is positive $\geq 1$, when $E$ is real $< \gamma$. For $E = \nu_k > \gamma$
but just above the cut along the real axis, we have from (2.23)
\begin{equation}
\text{Im} h_\infty (\nu_k + i o+) = i \left({g^2\over 4 \pi} \right) k w_k^2 ~.
\end{equation}
Thus, on the second sheet near and below the cut (2.25), there is a zero of $h_\infty (E)$,
which corresponds to the $V$-resonance. It can be shown that the phase shift $\delta$
for $Nc$ scattering (2.18) is related to $h_\infty (\nu_k + io+)$ and its complex
conjugate by
\begin{equation}
e^{-2i\delta} = { h_\infty (\nu_k + io+) \over [ h_\infty (\nu_k + io+)]^*}
\end{equation}

\section*{\Large \sf 3.  {\it Na} Sector (General Discussion)}
\setcounter{section}{3} \setcounter{equation}{0}
As discussed in the Introduction, assume the idealized case that there
does exist a fundamental spin $0$ field $\phi$ which is the origin of
masses of spin nonzero particles.
In any physical process, there are bound to be effective couplings
between $\phi$ and some of
its higher power products, such as
\begin{equation*}
\phi^2, ~\phi^3,~\cdots~,~\phi^n, ~\cdots~~.
\end{equation*}
Thus, the physical Higgs channel becomes connected to not only a complex
pole, but also to a cut in the complex energy plane, or other more complicated analytical structure.

In this section, the $Na$ channel that we shall analyze represents a highly idealized model of "Higgs" as
a composite. From reactions (2.10) and (2.11), we see that a state vector $|~\rangle$ in the
$Na$ sector must also have components in $Vb$
and $Nbc$ channels as well. Thus for $R$ finite, we may write
\begin{equation}
|~\rangle = \left[ \sum_k \psi(k) a^\dagger_k N^\dagger + \sum_p \phi(p) b^\dagger_p V^\dagger
+ \sum_{p,q} \chi(p,q) b^\dagger_p c^\dagger_q N^\dagger \right] |0~\rangle~~~.
\end{equation}

From
\begin{equation}
H|~\rangle= E|~\rangle~~~~,
\end{equation}
we find
\begin{equation}
(E-\lambda_k) \psi(k) = f U_k \sum_p V_p \phi(p)~~~,
\end{equation}
\begin{equation}
(E-m_0-\mu_p )\phi(p) = g \sum_q W_q \chi (p,q)+ f V_p \sum_k U_k \psi(k)
\end{equation}
and
\begin{equation}
(E-\mu_p-\nu_q )\chi(p,q) = g W_q \phi(q)
\end{equation}
where $k$, $p$, $q$ are all given by (2.5) and $U_k$, $V_p$, $W_q$
are related to the $u_k$, $v_p$ and $w_q$ of (2.3) by
\begin{equation}
\begin{array}{ll}
U_k & = (4 \pi R \lambda_k)^{-\frac{1}{2}} k u_k \\
V_p & = (4 \pi R \mu_p)^{-\frac{1}{2}} p v_p
\end{array}
\end{equation}
and
\begin{equation*}
W_q = (4 \pi R \nu_q)^{-\frac{1}{2}} q w_q
\end{equation*}

Substituting (3.5) into (3.4), we have
\begin{equation}
(E-m_0-\mu_p)\phi(p) = g^2 \sum_q W_q^2 (E-\mu_p - \nu_q)^{-1}\phi(p)
+ fV_p\sum_k U_k \psi(k)
\end{equation}
and therefore
\begin{equation}
\phi(p) = [D_p (E)]^{-1} f V_p \sum_k U_k \psi(k)
\end{equation}
where
\begin{equation}
D_p(E) = E-m_0-\mu_p -  g^2 \sum_q W_q^2 (E-\mu_p - \nu_q)^{-1}~~.
\end{equation}
From (3.8), we also have
\begin{equation*}
\sum_p V_p \phi(p) = f \left[\sum_p {V^2_{p} \over D_p (E)}\right]
\sum_k U_k \psi(k) ~~.
\end{equation*}
Thus, (3.3) becomes
\begin{equation}
(E - \lambda_k) \psi(k) = f^2 U_k \left[\sum_p {V^2_{p} \over D_p (E)}\right]
\sum_{k'} U_{k'}  \psi(k')
\end{equation}
Multiplying both sides by $U_k/(E-\lambda_k)$ and summing over $k$,
we find that the eigenvalue $E$ satisfies
\begin{equation}
1 = f^2 F(E) \sum_p {V^2_p \over D_p(E)}
\end{equation}
in which
\begin{equation}
F(E) =  \sum_k U^2_k (E - \lambda_k)^{-1}~~~~.
\end{equation}

Next, we study the continuum limit. When $R \rightarrow \infty$, the sum
\begin{equation}
U_k \sum_p V_p = (4\pi R)^{-1} \sum_p (\lambda_k \mu_p)^{-\frac{1}{2}} k u_k p v_p
\end{equation}
becomes
\begin{equation}
(4\pi^2)^{-1} \int (\lambda_k \mu_p)^{-\frac{1}{2}} k u_k p v_p dp~~~.
\end{equation}
Thus, from (3.3) we have
\begin{equation}
(E-\lambda_k) \psi(k) = (4\pi^2)^{-1}fk u_k \int (\lambda_k \mu_p)^{-\frac{1}{2}} p v_p \phi(p) dp~~~.
\end{equation}
Likewise, (3.7) leads to
\begin{equation}
\begin{array}{ll}
(E-m_0-\mu_p) \phi(p) &=\ds (4\pi^2)^{-1}g^2 \phi(p) \int \nu^{-1}_q (E-\mu_p-\nu_q)^{-1} q^2 w^2_q dq\\[0.5cm]
&+~\ds (4\pi^2)^{-1} fp v_p \int (\mu_p \lambda_k)^{- \frac{1}{2}} k u_k \psi(k) dk~~~,
\end{array}
\end{equation}
which gives
\begin{equation}
\phi(p) = (4\pi^2 \mathscr{D}_p(E) )^{-1}f  p v_p \int (\mu_p \lambda_k )^{-\frac{1}{2}} k u_k \psi(k) dk
\end{equation}
where
\begin{equation}
\mathscr{D}_p(E) = E-m_0-\mu_p - (4\pi^2)^{-1} g^2  \int [ \nu_q (E-\mu_p-\nu_q)]^{-1} q^2 w^2_q dq
\end{equation}

In a collision of $Na$, in order to describe reactions (2.13) and (2.14), we write $\psi(k)$
and $\phi(p)$ as
\begin{equation}
\psi(k) = \delta (k-k_0) + \widetilde{\psi} (k)
\end{equation}
and
\begin{equation}
\phi(p) = \widetilde{\phi} (p)
\end{equation}
in which $\widetilde{\psi}(k)$ and $\widetilde{\phi}(p)$ denote the scattered wave amplitudes. Thus,
(3.15) remains valid if we replace $\psi, ~\phi$ simply by $ \widetilde{\psi}$ and $ \widetilde{\phi}$.
Hence
\begin{equation}
(E-\lambda_k) \widetilde{\psi}(k) = (4\pi^2)^{-1}fk u_k \int (\lambda_k \mu_p)^{-\frac{1}{2}} p v_p \widetilde{\phi}(p) dp~~~.
\end{equation}
On the other hand, (3.17) yields
\begin{equation}
\widetilde{\phi}(p) = (4\pi^2 \mathscr{D}_p(E) )^{-1}f  p v_p \left[(\mu_p\lambda_0)^{-\frac{1}{2}}k_0 u_0+
\int (\mu_p \lambda_k )^{-\frac{1}{2}} k u_k \widetilde{\psi}(k) dk\right]
\end{equation}
with
\begin{equation}
\lambda_0 = \lambda_k~~~~~~\text{and}~~~~u_0 = u_k~~~~~\text{at}~~~~k=k_0~~.
\end{equation}
Define
\begin{equation}
\mathscr{A} = (4\pi^2)^{-1}\int (v_p p / \mu^{\frac{1}{2}}_p )\widetilde{\phi}(p)dp ~~~,
\end{equation}
\begin{equation}
\mathscr{B} = (4\pi^2)^{-1}\int (u_k k / \lambda^{\frac{1}{2}}_k) \widetilde{\psi}(k)dk ~~~
\end{equation}
and
\begin{equation}
\mathscr{C} = (4\pi^2)^{-1} (u_0 k_0 / \lambda^{\frac{1}{2}}_0 )~~~~~~~.~~~~~~~
\end{equation}
Hence, (3.21) and (3.22) can be written as
\begin{equation}
\widetilde{\psi}(k)= f \mathscr{A} u_k k / \lambda^{\frac{1}{2}}_k (E-\lambda_k)
\end{equation}
and
\begin{equation}
\widetilde{\phi}(p)= f (\mathscr{B}+ \mathscr{C}) v_p p / \mu^{\frac{1}{2}}_p \mathscr{D}_p(E)
\end{equation}
with $\mathscr{D}_p(E)$ given by (3.18). In above expressions, all integrations over $p$ and $k$ extend
from $0$ to $\infty$.

Substituting (3.28) into (3.24), we find
\begin{equation}
I \equiv \mathscr{A} / (\mathscr{B}+ \mathscr{C})
\end{equation}
is given by
\begin{equation}
I = {\frac{f}{4\pi^2}}\int_0^\infty { (v^2_p p^2/\mu_p)dp\over E-m_0-\mu_p -\ds \frac{g^2}{4\pi^2} \int_0^\infty
{q^2 w^2_q dq\over \nu_q (E-\mu_p-\nu_q)}}~~.
\end{equation}
Likewise,
\begin{equation}
J \equiv \mathscr{B} / \mathscr{A}
\end{equation}
becomes
\begin{equation}
J = {\frac{f}{4\pi^2}}\int_0^\infty { (u^2_k k^2/\lambda_k)dk \over E-\lambda_k}~~.
\end{equation}
Thus
\begin{equation}
\mathscr{A} = { I \over 1-IJ} \mathscr{C}
\end{equation}
and
\begin{equation}
\mathscr{B} +\mathscr{C}= { 1 \over 1-IJ} \mathscr{C}
\end{equation}
From (3.26), (3.30) and (3.32), we have the explicit expressions for
$\mathscr{C},~~I$ and $J$. Hence  $\mathscr{A}$ and $\mathscr{B}$ are
also known. Equation (3.27) and (3.28) then give scattering amplitudes
$\widetilde{\psi}(k)$ and $\widetilde{\phi}(p)$.

\section*{\Large \sf 4.  {\it Na} Sector (Critical $f^2$)}
\setcounter{section}{4} \setcounter{equation}{0}
We shall show that when $R \rightarrow\infty$ and $f^2$ greater
than a critical strength $f^2_c$, there exists a bound state in
the $Na$ sector. Write the $R \rightarrow\infty$ limit of (3.11) as
\begin{equation}
1 = f^2 \mathscr{F}(E) \mathscr{G}(E)
\end{equation}
in which
\begin{equation}
\begin{array}{ll}
\mathscr{F}(E) &= \ds \lim_{R \rightarrow\infty} F(E)\\[0.5cm]
& = \ds (4\pi^2)^{-1} \int^\infty_0 [ k^2 u^2_k /\lambda_k (E - \lambda_k)] dk
\end{array}
\end{equation}
with $u_k$, $\lambda_k$ given by (2.3) and (2.4). The function  $\mathscr{G}(E)$ is similarly
related to the last summation in (3.11) by
\begin{equation}
\begin{array}{ll}
\mathscr{G}(E) &= \ds \lim_{R \rightarrow\infty} \sum_p {V^2_p \over D_p(E)}\\[0.5cm]
& = \ds (4\pi^2)^{-1} \int^\infty_0 [ p^2 v^2_p /\mu_p \mathscr{D}_p(E) ] dp
\end{array}
\end{equation}
where $\mathscr{D}_p(E)$ is given by (3.18),
Since $\mathscr{D}_p(E)$ is related to $h_\infty (E) $ of (2.23) by
\begin{equation}
\mathscr{D}_p(E) = h_\infty (E -\mu_p)~~~,
\end{equation}
we have from (2.24)
\begin{equation}
\mathscr{D}_p(\mu_p + \gamma ) = h_\infty (\gamma) < 0 ~~~.
\end{equation}
From (2.4),
\begin{equation}
\nu_q  = (q^2 + \gamma^2) > \gamma ~~~.
\end{equation}
Thus, (3.18) and (4.4) - (4.5) imply that $\mathscr{D}_p(E)$ and its derivative
\begin{equation}
\mathscr{D}'_p(E)={\partial \over \partial E}\mathscr{D}_p(E)
\end{equation}
are continuous and satisfy
\begin{equation}
\mathscr{D}_p(E) <0 ~~~~~\text{and} ~~~~ \mathscr{D}'_p(E) > 0
\end{equation}
over the range
\begin{equation}
E < \mu_p + \gamma ~~~~,
\end{equation}
which includes the range
\begin{equation}
E < \alpha
\end{equation}
in accordance with (2.4) and (2.16). Thus, both
$\mathscr{F}(E)$ and $\mathscr{G}(E)$ are negative, with negative derivatives;
their product is positive
and varies from $0$ to a finite value as $E$ increases from $-\infty$ to $\alpha$,
the mass of $a$-meson.

Define a critical $f^2$-coupling by
\begin{equation}
f^2_c = [\mathscr{F}(\alpha)\mathscr{G}(\alpha)]^{-1}
\end{equation}
It then follows that there exists a bound state energy $E_0$ in the $Na$ sector with
\begin{equation}
1=  f^2  \mathscr{F}(E_0)\mathscr{G}(E_0)
\end{equation}
provided
\begin{equation}
f^2 > f^2_c
\end{equation}
For $f^2 < f^2_c$, the state turns into a resonance with a complex $E_0$.
In this case the scattering amplitudes $\widetilde{\psi}(k)$ and
$\widetilde{\phi}(k)$ have besides the cuts given by (3.30) and (3.32), also a complex
pole at $E=E_0$.

It is of interest to note the difference between this pole in the $Na$ sector and the $V$-pole in the $Nc$ sector.
The $V$-pole becomes stable in the weak coupling limit when
$g^2 \rightarrow 0$, whereas in the $Na$ sector the boundstate $E_0$
becomes stable only in the strong coupling region when $f^2 > f^2_c$.

We note that when $g^2=0$, $\mathscr{G}(E)$ of (4.3) becomes
\begin{equation}
\mathscr{G}_0(E) = \ds (4\pi^2)^{-1} \int^\infty_0 \left[ p^2 v^2_p /\mu_p
(E-m_0-\mu_p) \right] dp~~~~.
\end{equation}
Correspondingly, (4.12) becomes
\begin{equation}
1= f^2 \mathscr{F}(E_0) \mathscr{G}_0(E_0)
\end{equation}
with the same $\mathscr{F}(E)$ of (4.2). Thus, the existence of the pole at
$E=E_0$ does not depend sensitively on $g^2$; instead, it is closely
related to the second (and higher) order attractive potential
between $Na$ due to the $f$-coupling transitions
\begin{equation}
Na \rightleftarrows Vb \rightleftarrows Na~~~.
\end{equation}
Its physical origin is quite different from the $V$-pole in the $Nc$
channel of (2.18).

\newpage
\section*{\Large \sf 5. Remarks}
\setcounter{section}{5} \setcounter{equation}{0}

Consider the case
\begin{equation}
f^2 < f^2_c~~~~.
\end{equation}
The composite $Vb$ is unstable, and may serve as a highly simplified model of
either the $\sigma$-meson or the Higgs boson. Besides the elastic process
(4.16), there is also the inelastic reaction
\begin{equation}
Na \rightleftarrows Vb \rightleftarrows Nbc~~~.
\end{equation}
In order to detect the composite $Vb$ as a resonance, we require in (4.12) the corresponding pole
at
\begin{equation}
E = E_0
\end{equation}
to be not too far from the real axis; hence $f^2$ cannot be too small. The amplitude for
the continuum background must then also be relatively large.

In any composite model, we may regard the amplitudes $Na$
and $Vb$ as the idealized representations of its low and high
frequency components of the same composite state vector.
A second order transition between these two components would always depress $Na$ and
elevate $Vb$, as in (4.16). A resonance thus formed would require a strong coupling,
and therefore also a large continuum background as in the model.
This could be the reason why the $\sigma$-meson
does not appear as a sharp resonance, and it might also be difficult to isolate
the Higgs boson resonance.

We wish to thank N. Christ and E. Ponton for discussions.
\newpage
\section*{\Large \sf Appendix}
\renewcommand{\thesection}{\Alph{section}}
\setcounter{section}{1} \setcounter{equation}{0}
In the special case when
\begin{equation}
w^2_k = k/\nu^3_k~~~,
\end{equation}
the integral
\begin{equation}
F_\gamma(E) = \int^\infty_0 {k^2w^2_k \over 4\pi^2\nu_k} {dk\over (E-\nu_k)}
\end{equation}
in (2.23) is given by
\begin{equation}
F_\gamma(E) = {1\over 4\pi^2\gamma} \left[ {1\over z^3} (1-z^2)\text{ln} {1\over 1-z} -
{1\over 2z^2} (z+2) \right]
\end{equation}
with
\begin{equation}
z = E/\gamma~~~~.
\end{equation}
At $E=\nu_k + i 0+$, we have
\begin{equation}
\text{Im} F_\gamma(E) = {-i\over 4\pi\gamma} \left[ {1\over z^3} (z^2-1)\right]
\end{equation}
in agreement with (2.27).

\newpage

\section*{\Large \sf References}

{\normalsize \sf

\noindent [1] J. Schwinger, Ann. Phys. (N.Y.) {\bf 2}, 407(1957). J. C. Polkinghorne,
Nuovo Cimento {\bf 8}, 179(1958); {\bf 8}, 781(1958).
M. Gell-mann and M. Levy, Nuovo Cimento {\bf 16}, 705(1960).\\
~[2] C. Amsler {\it et al.} (Particle Data G Group), Physics Letters {\bf B667}, 1(2008) and
2009 partial update for the 2010 edition listed
$f_0(600)$ of mass $400-1200~Mev$ and full width
$600-1000~Mev$, which closely resembles the appearance of a composite, similar to $Na
\rightarrow Nbc$ of (2.14), that will be analyzed in this paper.\\
~[3]  P. W. Higgs, Phys. Rev. Lett  {\bf 12}, 132(1964); Phys. Rev.  {\bf 145}, 1156(1966).\\
~[4]  S. Weinberg, Phys. Rev. {\bf D13}, 974(1976).\\
~[5]  L. Susskind, Phys. Rev. {\bf D20}, 2619(1979).\\
~[6]  C. T. Hill and E. H. Simmons, Phys. Report. {\bf 381}, 235(2003).\\
~[7]  R. Contino, Y. Nomura and A. Pomarol, Nucl. Phys. {\bf B671}, 148(2003).\\
~[8]  K. Agashe, R. Contino and A. Pomarol, Nucl. Phys. {\bf B717}, 165(2005).\\
~[9]  T. D. Lee, Phys. Rev.  {\bf 95}, 1329(1954).\\
~[10]  G. K\"{a}llen and W. Pauli, Dan. Mat. Fys. Medd.  {\bf 30}, 7(1955).\\
~[11]  S. Weinberg, Phys. Rev.  {\bf 102}, 285(1955).\\
~[12]  V. Glaser and G. K\"{a}llen, Nucl. Phys.  {\bf 2}, 706(1956/57).\\
~[13]  G. K\"{a}llen, Renormalization Theory in "Lectures in Theoretical Physics",
W. A. Benjamin, New York(1962), Vol. 1, pp. 169-256.\\
~[14]  W. Heisenberg, Proceedings of the Symposium on Basic Questions in Elementary Particle Physics,
June 8-18, 1971. Munich: Max-Planck Institute for Physics and Astrophysics: 1-10.\\
~[15]  T. D. Lee and G. C. Wick,  Nucl. Phys.  {\bf B9}, 209-243(1969); Nucl. Phys.  {\bf B10}, 1-10(1969).

}

\end{document}